# Yuan: Research on the Concept of Digital World Analogue Scientific Infrastructure and Science Popularization Communication Based on Suzhou Gardens Pattern

Zhang Lvyang*‡, Lu Wen*‡, Zhao Yang*‡, Li Jiaqi*‡, Zhai Lidong‡

*School of Cyber Security, University of Chinese Academy of Sciences, Beijing, China

‡Institute of Information Engineering Chinese Academy of Sciences, Beijing, China

Email: zhanglvyang@iie.ac.cn

*Abstract*-In the current digital era, high security relies significantly on advanced concepts such as native security. However, the design and implementation of these concepts face challenges in enterprises and organizations. Leveraging advancements in Large Language Models (LLMs), we draw inspiration from the design principles of Suzhou Gardens, a UNESCO World Heritage site. By examining its core features, which align closely with those of the AI world simulator Sora, we extract three concurrent concepts to enhance the security of future digital infrastructures.We propose three guiding principles to facilitate the preliminary construction of the "Space Spider," a hyper-large scientific infrastructure. These principles will steer the development of the "Yuan" digital garden, establishing a "Chinese Series" focused on the construction pathways of the Yuan AI world simulator. The initial pilot of Yuan is expected to generalize various hyper-large scientific infrastructure scenarios, ultimately expanding into numerous high-security digital applications.Through the design concept of Suzhou Gardens, we aim to promote science communication and talent training in the field of cybersecurity. With the support of Yuan, we intend to extend our efforts to various digital construction domains. This initiative is poised to contribute significantly to the future of digital world simulators, emphasizing the integration of hyper-large scientific infrastructure with science communication and research dissemination.

## I. Introduction

In the current digital era, the threat situation and risk vectors in the field of cybersecurity on a global scale are becoming increasingly complex, and the limitations of mechanisms and methods of traditional cyber attack-defence simulation are gradually highlighted, exacerbating the current cyber security challenges. The global economy is undergoing a digital transformation of thousands of industries, and its digital security observation perspective has also undergone profound changes, with the accompanying security model of traditional cybersecurity iterating to endogenous security model, and even to nascent security model. For instance, the EU Cyber Resilience Act (hereinafter referred to as "CRA"), which will come into force on 1 July 2024, introducing a legal framework that requires all products with digital elements sold on the EU market to be secure from design to the entire life cycle, which fully embodies the design concept of native security.

In fact, Microsoft's Secure Development Lifecycle (SDL) embeds comprehensive security requirements, technology-specific tools, and required processes into the development and operation of all software products, and DevSecOps defines security as the responsibility of all members of the entire IT team, including development, testing, operations, and security teams, from development to operations. The deployment of native security concepts helps to detect problems early enough, and the large infrastructure software mentioned above has already deployed native security concepts in its products in advance. However, the remote code execution vulnerability of log4j2 [cve-2021-44228], which was reported on the 8th of December 2021, meanwhile, exposed the limitations of the current deployment of native security concepts. Current software development is evolving from the open source era to the low-code and zero-code era, so the native security of most componentized open source components cannot be effectively guaranteed by deploying them in advance.

Therefore, the deployment of nascent security concepts in the digital security era needs to take a different approach. Compared with traditional cyber security, digital security has three differences: firstly, it is at a different evolutionary level; secondly, it is at a different observation dimension; and thirdly, it is at a different weighting relationship of the main research object [1]. Both the Microsoft SDL development and the Business Innovation Acts in the US and the EU deploy security front-loading mentioned above, are deploying native security design concepts from a regulatory perspective, which means they rarely consider security design from a practical business perspective.

With the integration of AI and the digital economy, the path of industrial digital development is becoming increasingly variable with the continuous improvement of the intelligent deduction capability of AI LLM. Different from the previous "sensitive state" and "steady state" mode, the current industrial digitalization is more inclined to a "hybrid state" development mode, and its connotation and concept have also shifted sharply.

In recent years, AI and virtual simulation fusion technology, represented by digital twin technology, has shown a strong potential for application. Digital twin not only provides support for cutting-edge technologies such as metaverse, but also plays a significant role in the digital transformation of various industries, meanwhile providing a technical basis for the development of digital "hybrid state". The migration of many entity systems from the physical world to the digital world has provided the digital world with a wealth of possibilities and multiple scenarios: for instance, the rapid generation of a 1-minute high-quality video in February 2024 using the OpenAI Sora video generation model technology has further accelerated the rapid construction of multiple scenarios for digital twin. Owning a sufficiently huge number of scenarios is one of the prerequisites for the high-speed advancement of digital twin and metaverse technologies.

Suzhou Gardens, as a representative of Jiangnan Gardens, is listed in the World Cultural Heritage, which is the condensation and crystallization of the aesthetics of classical Chinese garden art, combining such aesthetical interests with digital twin technology to innovate the model of cybersecurity science popularization communication dissemination can inject more cultural connotations into the science popularization communication dissemination scenarios in the field of cyber security.

Therefore, this paper proposes a method of research on the concept of digital world simulation scientific device and science popularization communication dissemination based on the Suzhou Gardens pattern, using the innovative application of AI digital twin technology to construct diversified scenarios to promote the research on science popularization communication dissemination in the field of cybersecurity, and exploring the construction of a representative of the AI world simulator of the "Chinese series", named "Yuan". Through the empowerment of the Suzhou Gardens design concept, the pan-scene digital twin technology provides a new path and a reference model for the dissemination of cyber security science popularization communication dissemination and talent training.

## II. Analysis of Suzhou Garden Design Concept and World Simulator Integration Construction Scenario

By demonstrating the application of digital twin technology in different fields, scientific and technological knowledge can be more vividly presented to the public, and the public's interest and understanding of digital technology can be improved consequently. The automated generation of digital twin scenes based on the design concept of Suzhou Gardens can help generate widespread and diverse digital twin scenes of cyber security, and build an innovative and interesting testing ground for cyber security science popularization communication dissemination and technology dissemination. For example: by simulating digital security scenarios, it can help students understand cyber threats and digital security challenges, and cultivate their sensitivity and response ability to digital security; by constructing a multi-scenario digital twin approach experimental environment, it can improve the experimental hardware and software experimental conditions of talent training with lower cost and higher efficiency. The application of digital twin technology in science popularization communication dissemination and talent cultivation can promote cyber security education from the traditional "multi-modal" era into the "pan-scene" era, so that the science popularization communication dissemination of cyber security has become closer to the real education condition, more interesting and fascinating.

In the current era of digital security, against the three essential differences between digital security and traditional security mentioned in the introduction, the feasibility and integrability of carrying out cybersecurity science communication dissemination based on the design concept of Suzhou Garden and the world simulator represented by Sora to jointly empower multiple scenarios, the main reasons include the following 3 points. Firstly, Open AI adopts diffusion transformer to solve the problem of spacetime interlacing and unification. On 15-February 2024, Sora officially announced the release of AI-generated 1-minute high-quality video, which has proved that this problem has been solved. Likewise, Suzhou Gardens are solid ensemble, they have properties of long-term existence and full sensory coverage, and their embodiment is the comprehensive superposition of the established human perception system, beyond the current video, audio or other modalities of expressional and experience range: video mainly embodies the visual, auditory, but cannot embody other main human senses such as taste, touch, smell, etc., hence the design concept Suzhou Gardens and Sora and other scenes as the main carrier of the AI. Therefore, the combination of the Suzhou Gardens-based concept and AI world simulators presented by Sora, which are mainly based on scenarios, is very promising, and such simulators can provide more diversified and numerable scenarios for cyber security talent training and science popularization communication dissemination.

Secondly, Suzhou Gardens contain a wealth of non-human organisms, such as animals, plants, insects, etc. Based on the design concept of Suzhou Gardens, the world simulator can build a pan-scenario environment that is more similar to the "ecosystem" and beyond human society, which represents a more cutting-edged direction of development and a further-reaching dimension of evolution for AI world simulators, i.e., from person-centric to humankind-centric. This represents a tremendous milestone leap from person-centred to humankind-centred till everything-centred.

Thirdly, one of the core ideas of Sora's video generation process is patchify, and the patches are trained in the latent space to achieve compression, feature extraction, data dimensionality reduction, thus making the cost reduce sharply through the standardization and encapsulation. The characteristics of Suzhou Gardens are highly compatible with this: Suzhou Gardens are characterized by "emptiness", which does not mean nothing at all, but as the Buddhist saying that "a flower is a world, a leaf is a bodhi", which is also the essential characteristic of metaverse. This is also the original meaning of the Chinese word "Yuan" in the metaverse. Suzhou gardens are fixed architectural complexes, and under normal conditions, they do not change drastically over time, so "emptiness" is meaningful to people and everything. Therefore, "empty" and "yuan" can fully match the core meaning of "patch", i.e., standardizing the digital security as "Yuan".

We define the multiple scenes generated by AI digital twin as "scenario". "One thousand readers have a thousand Hamlets", which is the effect of multi-scenario communication brought by multiple "scenarios". As we all know, the scenario of "Peony Pavilion", whose plot originates from " Romance of the Western Chamber ", the scene is derived from Suzhou Gardens, and the characters are Du Liniang and Liu Mengmei, is a complete story, with 6 elements of theatre creation, so the scenario " Peony Pavilion ", relying on Suzhou Gardens, can easily continue to automate the production of " Peony Pavilion 2", " Peony Pavilion 3", etc., with the help of AI models such as ChatGPT and Sora. etc., by keeping giving the initial scene input of the derivative work; and using other tools to automate the character dialogue to achieve a variety of languages dubbing work, it is also not difficult for AI tools. It, obviously ,is easy to realize that " Peony Pavilion " is one of the typical "scenarios" based on the design concept of Suzhou gardens, in addition to the "Peony Pavilion" play written by Tang Xianzu as well as other literary masterpieces, there are many similar "scenarios" that took place in Suzhou Gardens.

III. Path Analysis of High Security Digital Infrastructure in the Future

A typical case of building future high-security digital infrastructure based on the design concept of Suzhou Gardens is the construction of digital large science infrastructures. The traditional big scientific devices are typical steady state scientific infrastructure, while in the digital world, the construction of its target form and the evolution of its characteristics, in terms of cost, function and impact, will typically reflect the development of the "mixed state" attributes. Therefore, we have selected a Hyper Large Scientific Infrastructure, the Space Spider Digital Industrial Experimental Bed, as a typical research object. The Space Spider Hyper Large Scientific Infrastructure has 12 major characteristics, around which the Space Spider Digital Industrial Experimental Bed can be built. Of course, it can also be built according to other ideas of the digital industry experimental bed, although the construction of its capabilities, paths are different, but different routes come to the same destination, they all can be used as a digital industry experimental bed of the hyper large scientific infrastructure [2]. There are three reasons why we carry out the future high-security digital infrastructure based on the design concept of Suzhou Gardens: Firstly, the construction of Suzhou Gardens has a certain threshold, but the threshold is not too high, bureaucrats, squires, businessmen, scholars and other people with different identities can build gardens of different sizes and styles as their own mansions; this point is matched with the characteristics of digital transformation of thousands of industries with their own apparent styles. Secondly, the design concept and construction process of Suzhou Gardens implies the "5-dimensional" feature of "express motion through stillness ": the spatial metrics include three dimensions, x, y and z, together with the fourth dimension--time, and the fifth dimension--illusion, which together constitute the five design dimensions of Suzhou Gardens. Suzhou Gardens will produce different scenes according to the time lapse, each moment of a certain Suzhou Garden is a new independent Suzhou Garden, this is the design concept of the time dimension; each visitor will have different subjective feelings in the same place when they revisit Suzhou Garden, that is, each person will produce a host of illusions in the "Suzhou Garden", and different people will have different experiences to make the " illusory Suzhou Garden". The different experiences of different people will increase the number of illusions to infinity. These five dimensions of the design concept make the core features of Suzhou Gardens fit perfectly with the features of the digital world. The schematic diagram of Suzhou Gardens design concept and digital world high security deployment is shown in Figure 1.

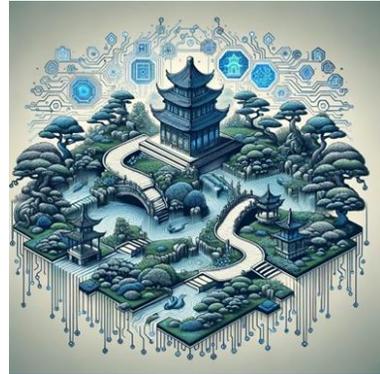

Fig. 1 Schematic diagram of Suzhou Gardens design concept and digital world high security deployment

The history of Suzhou Gardens construction is over hundreds of years. The construction and evolution of Suzhou Gardens reflect the cultural and aesthetic alters in different historical periods, and have a thick historical and cultural accumulation. In addition, Suzhou Gardens incorporate a vast host of elements of literature, painting and poetry, reflecting the deep innermost cultural content of traditional Chinese culture.

The connotation of Suzhou Gardens is contained in the traditional Chinese character "園". In the book "Jiangnan Garden Journal" written by Mr. Tong Jun, the chapter of "Gardening" highly summarizes the elements, techniques and realm of classical Chinese gardens. Mr. Tong broke down the Chinese character "園" to get the elements of a garden: courtyard walls, buildings, pools, flowers, trees, rocks and mountains. The outer "囗" is the enclosing wall. The inner character "土" resembles the plane of a house and can represent a pavilion. The inner character "口" is centred on the pool. The inner character "从" in front resembles a stone or a tree. The gardening techniques summed up as " virtualization and reality echoed, huge and tiny contrast, high and low layout proportionality", summed up the realm that gardening "three realms": "sparse and dense, the curves and twists and turns, there is a scene in front of your eyes" [3].

Through elaborate layout, traditional garden elements unified in a limited space in Suzhou Gardens, such as courtyard walls, buildings, pools, flowers, trees, rocks and other elements in the Suzhou Gardens has been exquisite, compact and timeless expressed. The Humble Administrator's Garden, the Lingering Garden, and the Master of Nets Garden are even designed to incorporate the sentiments of the literati, making the gardens not only landscapes, but also a concentrated exposition and high concentration of the aesthetic interests of nationalism; and the generalization and concentration are precisely the essence of Suzhou Gardens' design concepts.

Similarly, based on the design concept of Suzhou Gardens, the high-security digital infrastructure in the future digital world should also have the corresponding design elements of generalization, i.e. S-C-D-E elements, which are ordered according to the consideration of construction principles, namely scientific paradigm S (scientification), C (connotation), D (design),and E(evaluative). Its schematic diagram of the S-C-D-E features is shown in Figure 2.

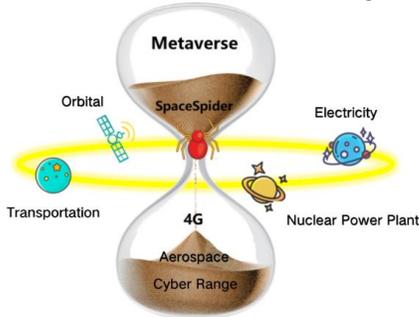

Fig. 2. Hyper Large Scientific Infrastructure: Spatio-temporal transcendence

With the design elements of S-C-D-E, it is possible to condense a variety of forms of "illusion", so that the digital construction of thousands of trades and industries in a parallel universe similar to the film "Spider-Man: Parallel Universe" in the parallel universe of the space to be entirely presented.

IV. Deployment and Implementation of High Security Concepts in the Digital World: The Construction of AI Twin Scenarios for Digital Security in the "Chinese Series" Represented by "Yuan".

The high security of the digital world should embody three principles of the deployment of native digital security , which can be generalized as 3 principles: **first**, attaching equal emphasis on basic and advanced defenses; **second**, deploying self-defense mode and escort mode strategies simultaneously; **third,** promoting the co-growth of security strategies and industry situations. These 3 principles can correspond to 3 essential differences between digital security and traditional cyber security: different evolutionary levels at which they are located, different observation dimensions, and different weighting relationships of the main research objects. Specific high-security protection measures can be carried out in the following spheres: anti-APT attacks, active defense, asset maps and full-cycle protection, situational awareness, target range intelligent deduction, industrial intelligence and early warning, etc.[4], and its evolutionary level and implementation stage can be summarized as CIID [5]. Schematic diagram of the Space Spider Garden based on the design concept of Suzhou Gardens is shown in Figure 3.

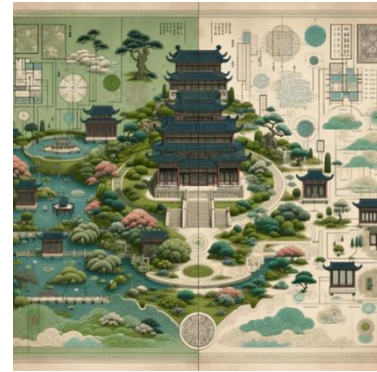

Fig. 3 Schematic diagram of the Space Spider Gardens based on the design concept of Suzhou Gardens

Using the design concept of integrating Suzhou Gardens with the digital world, we use the AI video generation tool to build a digital " Space Spider Garden" named "Yuan". Yuan" transfers the design concept of Suzhou Gardens with one motion and one stillness to the field of digital security twin scene generation, and its evolutionary level and implementation stage can be divided into "five births and three quantities" [6]. The construction of "Yuan" digital security twin scene can be understood as each part of the physical world of the Space Spider Hyper Large Scientific Infrastructure corresponds to an element or node in the digital world. Each plant in "Yuan" may represent a node in the network, ponds and rivers may symbolize the data flow in the network, and buildings may correspond to servers or databases. "Yuan" abstracts the composition of the physical world's Hyper Large Scientific Infrastructure into a data model, forming a digital virtual environment that can be interacted with, simulated, and analyzed, which can be used to design a security resilience system in the field of digital security, capable of resisting external threats while maintaining the robust operation of the internal elements. If a part of "Yuan" is subject to a security threat: for example, if a kiosk representing a database is attacked by a ransomware attack, the system will simulate the scope of the impact and possible knock-on effects in real time. The system, in real time, will simulate the scope of the impact and possible chain reactions, test different defense strategies and optimize security configurations, identify potential risks and predict possible attack paths, achieving the goal of fortifying the high-security twin scenarios of the Suzhou Gardens' concept-enabled Space Spider Industrial Digital Experiment Bed. The content of "Yuan" digital world simulation scientific infrastructure is schematically shown in Figure 4-Figure 7.

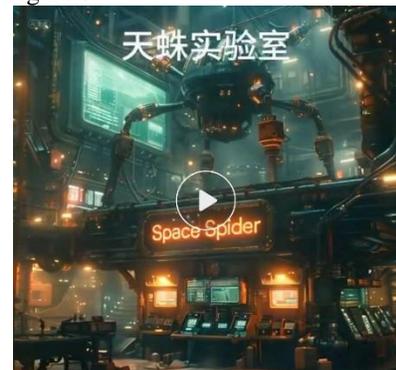

Fig. 4 Schematic representation of the contents of the

"Yuan" digital world analogue scientific installation.

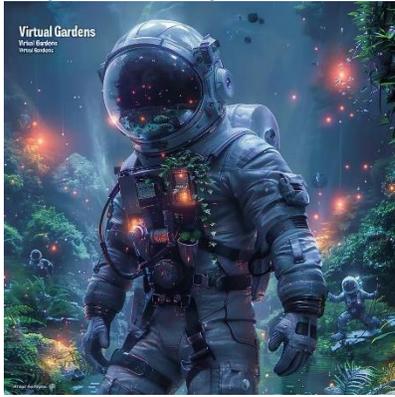

Fig. 5 Schematic representation of the contents of the "Yuan" digital world analogue scientific installation.

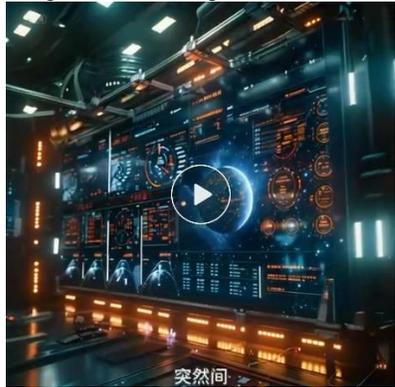

Fig. 6 Schematic representation of the contents of the "Yuan" digital world analogue scientific installation.

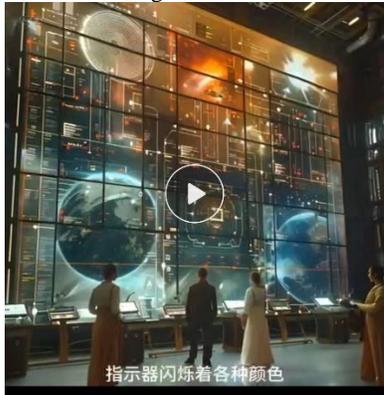

Fig.7 Schematic representation of the contents of the "Yuan" digital world analogue scientific installation.

"Yuan" is not only an attempt of AI digital twin technology to empower the digital security science popularization communication dissemination and the deployment and implementation of the concept of high security in the digital world, but also a significant exploration of the construction of digital world simulators for the digital construction of China's industries in the future, and it will become a typical "Chinese Series" digital world simulator comparable to the German or American ones, thus building a cluster of AI digital twin simulators across the globe.

## V. Conclusion and Prospect

Relying on the design concept of Suzhou Gardens to carry out the pilot test of the future high-security digital infrastructure - "Yuan", which represents that cutting-edge technologies such as digital twin and metaverse are providing the construction of security resilience capacity of the digital world from multi-modal to pan-scenario, which will support for security resilience capacity building in the digital world. Through the "Yuan" digital world simulator, the construction practice of the Space Spider Hyper Large Scientific Infrastructure and the exploration of science dissemination will lead to the digital transformation of more industries and provide more reliable security reinforcement strategies and risk prevention means.

In the future, we will continue to deepen the automated generation of digital security twin scenarios and accelerate the process of multi-modal and pan-scenario leapfrogging through the integration of Suzhou Gardens-based design concepts with the training of future cyber security talents and science popularization communication dissemination. We plan to integrate different scenarios and playful moods into the digital security field, and incorporate pan-scenario functions into the "Space Spider" experimental bed, so that more enterprises and organizations can attain cost reductions and efficiency gains by digital means. For instance, science fiction novel writers can carry out brainstorming in "Yuan" to help them plan their work and stimulate their ideas; film directors with small production costs can carry out digital twin simulation filming in "Yuan". By using Sora, film directors may direct low-cost documentaries on mysterious celebrities who passed away for a long period, such as Lin Huiyin, or make digital twin films on expressing the emotions between animals or aliens.

We look forward to attracting more creators and professionals in the field to participate in the "Yuan" World Simulator, and to jointly promote innovation and development in the field of digital security. We can look forward to more gifted litterateurs such as "Tang Xianzu" in the future to draw abundant inspiration from this digital "Yuan" and create even more outstanding masterpieces.